\documentclass[12pt]{article}
\usepackage{amsfonts,amsmath,amsthm}
\usepackage{graphicx}

\def\be{\begin{equation}}
\def\ee{\end{equation}}
\def\ba{\begin{array}}
\def\ea{\end{array}}

\def\bee{\begin{eqnarray}}
\def\eee{\end{eqnarray}}

\def\dis{\displaystyle}

\begin{document}
\title {The Basic Reproductive Number of Ebola and the Effects 
of Public Health Measures: The Cases of Congo and Uganda}

\author{G.\ Chowell$^{1,4}$, N. \ W. Hengartner$^{2}$,
  C. Castillo-Chavez$^{1,4}$, \\ P. \ W. Fenimore$^{1}$, J. \ M. Hyman$^{3}$ \\
\footnotesize $^{1}$ Center for Nonlinear Studies (MS B258) \\
\footnotesize $^{2}$ Statistical Science (MS F600) \\
\footnotesize $^{3}$ Mathematical Modeling and Analysis (MS B284)  \\
\footnotesize Los Alamos National Laboratory \\
\footnotesize Los Alamos, NM 87545 \\
\footnotesize $^{4}$ Department of Biological Statistics and
Computational Biology\\
\footnotesize Cornell University \\
\footnotesize Warren Hall, Ithaca, NY 14853\\
\footnotesize LA-UR-03-8189 \\
\footnotesize (Abstract: 174 words, Text: 3,295 words)
}

\date{}

\maketitle

\begin{abstract}

Despite improved control measures, Ebola remains a serious 
public health risk in African regions where recurrent outbreaks have been observed 
since the initial epidemic in $1976$. Using epidemic modeling and data from two 
well-documented Ebola outbreaks (Congo $1995$ and Uganda $2000$), we
estimate the number of secondary cases generated by an index case
in the absence of control interventions ($R_0$). Our estimate of 
$R_0$ is $1.83$ (SD $0.06$) for Congo (1995) and
$1.34$ (SD $0.03$) for Uganda (2000). We model the course of the outbreaks via
an SEIR (susceptible-exposed-infectious-removed) epidemic model that
includes a smooth transition in the transmission rate after control interventions are put in
place. We perform an uncertainty analysis of the basic reproductive number $R_0$ to quantify its sensitivity to other disease-related parameters. We also analyze the sensitivity of the final epidemic size to the time interventions begin and provide a distribution for the final epidemic size. The control measures implemented during these two outbreaks (including education and contact tracing followed by quarantine) reduce the final epidemic
size by a factor of $2$ relative the final size with a two-week delay in their implementation.
\end{abstract}

\section{Introduction}

\indent Ebola hemorrhagic fever is a highly infectious and lethal disease named after a
river in the Democratic Republic of the Congo (formerly Zaire) where it 
was first identified in 1976 \cite{CDC1}. Twelve outbreaks of Ebola
have been reported in Congo, Sudan, Gabon, and Uganda as of September 14, 2003
\cite{CDC2,WHO0}. Two different strains of the Ebola virus (Ebola-Zaire and the Ebola-Sudan) 
have been reported in those regions. Despite extensive search,
the reservoir of the Ebola virus has not yet been identified
\cite{Breman2, Leirs1}. Ebola is transmitted by physical contact with
body fluids, secretions, tissues or semen from infected persons \cite{CDC1, WHO1}. Nosocomial
transmission (transmission from patients within hospital settings) has 
been typical as patients are often treated by
unprepared hospital personnel (barrier nursing techniques need to be observed). 
Individuals exposed to the virus who become infectious do so after a mean incubation 
period of $6.3$ days ($1-21$ days) \cite{Breman1}. Ebola is
characterized by initial flu-like symptoms
which rapidly progress to vomiting, diarrhea, rash, and internal and external bleeding.
Infected individuals receive limited care as no specific treatment
or vaccine exists. Most infected persons die within $10$ days of 
their initial infection \cite{Nature1} ($50\%-90\%$ mortality \cite{WHO1}).\\

\begin{figure}[h*]
   \begin{center}
   \scalebox{0.5}{\includegraphics{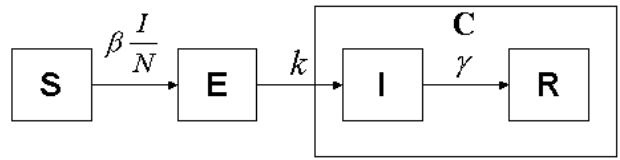}}
   \end{center}
   \caption{A schematic representation of the flow of individuals
   between epidemiological classes. $\beta \frac{I}{N}$ is the
   transmission rate to susceptibles $S$ from $I$; $E$ is the class
   of infected (not yet infectious) individuals; $k$ is the rate at
   which $E$-individuals move to the symptomatic and infectious class
   $I$; Infectious individuals ($I$) either die or recover at rate
   $\gamma$. $C$ is not an epidemiological state but keeps track of
   the cumulative number of cases after the time of onset of symptoms.}
\label{myfig0}
\end {figure}

\noindent Using a simple SEIR (susceptible-exposed-infectious-removed) 
epidemic model (Figure \ref{myfig0}) and data from two 
well-documented Ebola outbreaks (Congo $1995$ and Uganda $2000$), we
estimate the number of secondary cases generated by an index case
in the absence of control interventions ($R_0$). Our estimates of
$R_0$ are $1.83$ (SD $0.06$) for Congo (1995) and
$1.34$ (SD $0.03$) for Uganda (2000). We model the course of the outbreaks via
an SEIR epidemic model that includes a smooth transition in the transmission rate after
control interventions are put in place. We also perform an uncertainty
analysis on the basic reproductive number $R_0$ to
account for its sensitivity to disease-related parameters and analyze the
model sensitivity of the final epidemic size to the time at which
interventions begin. We provide a distribution for the final epidemic size. A 
two-week delay in implementing public health measures results in an
approximated doubling of the final epidemic size.
\section{Methods}

We fit data from Ebola hemorrhagic fever outbreaks in Congo (1995) and Uganda (2000)
to a simple deterministic (continuous
time) SEIR epidemic model (Figure \ref{myfig0}). The least-squares fit
of the model provides estimates for the epidemic parameters. The fitted model can then be used to estimate 
the basic reproductive number $R_0$ and quantify the impact of intervention measures on the
transmission rate of the disease. Interpreting the fitted model as an
expected value of a Markov process, we use multiple stochastic
realizations of the epidemic to estimate a distribution for the final
epidemic size. We also study the sensitivity
of the final epidemic size to the timing of interventions 
and perform an uncertainty analysis on $R_0$ to account for
the high variability in disease-related parameters in our model.

\subsection{Epidemic Models}
Individuals are assumed to be in one of the following epidemiological
states (Figure \ref{myfig0}): susceptibles (at risk of contracting the disease),
exposed (infected but not yet infectious), infectives (capable of
transmitting the disease), and removed (those who recover or die from the
disease). 

\subsubsection{Differential Equation Model}
Susceptible individuals in class $S$ in contact with the
virus enter the exposed class $E$ at the per-capita rate $\beta
I/N$, where $\beta$ is transmission rate per
person per day, $N$ is the total effective population size, 
and $I/N$ is the probability that a contact is made with a infectious 
individual (i.e. uniform mixing is assumed). 
Exposed individuals undergo an average incubation period (assumed
asymptomatic and uninfectious) of $1/k$ days before progressing to the 
infectious class $I$. Infectious individuals move to the $R$-class
(death or recovered) at the per-capita rate $\gamma$ (see Figure
\ref{myfig0}). The above transmission process is modeled by the following system of
nonlinear ordinary differential equations \cite{AM, BC}:\\
\begin{equation}
\label{eqn1}
\begin{array}{rcl}
      {\dis \dot{S}(t)}&=& -\beta S(t) I(t)/N\\
      {\dis \dot{E}(t)}&=& \beta S(t) I(t)/N -k E(t)\\ 
      {\dis \dot{I}(t)}&=& k E(t) - \gamma I(t)\\
      {\dis \dot{R}(t)}&=& \gamma I(t)\\
      {\dis \dot{C}(t)}&=& k E(t),\\
\end{array}
\end{equation}

\noindent where $S(t)$, $E(t)$, $I(t)$, and $R(t)$ denote the number of
      susceptible, exposed, infectious, and removed individuals at
      time $t$ (the dot denotes time derivatives). $C(t)$ is not an epidemiological 
      state but serves to keep track of
      the cumulative number of Ebola cases from the time of onset of symptoms.\\

\subsubsection{Markov Chain Model}
\noindent The analogous stochastic model (continuous time Markov
chain) is constructed by considering three events: \textit{exposure}, \textit{infection} and
\textit{removal}. The transition rates are defined as:\\

\begin {tabular}{l c l}
\hline
Event & Effect & Transition rate\\
\hline
Exposure & (S, E, I, R) \ $\rightarrow$ \ (S-1, E+1, I, R) & $\beta(t) S I/N$ \\
Infection & (S, E, I, R) \ $\rightarrow$ \ (S, E-1, I+1, R) & $k E$ \\
Removal & (S, E, I, R) \ $\rightarrow$ \ (S, E, I-1, R+1) & $\gamma I$ \\
\hline
\\
\\
\end{tabular}

\noindent The event times $0 < T_1 < T_2 < ...$ at which an
individual moves from one state to another are modeled as a renewal
process with increments distributed exponentially, \\

$$P(T_k - T_{k-1} > t  | T_j, j \le k-1) = e^{-t \mu(T_{k-1})}$$

\noindent where $\mu(T_{k-1}) = (\beta(T_{k-1}) S(T_{k-1})I(T_{k-1})/N
+ kE(T_{k-1}) +\gamma I(T_{k-1}))^{-1}$.\\

\noindent The final epidemic size is $Z=C(T)$ where
$T= min\{t>0, E(t)+I(t)=0 \}$, and its empirical distribution can be
computed via Monte Carlo simulations \cite{Renshaw1}.\\

\subsection{The Transmission Rate and the Impact of Interventions}

\noindent The intervention strategies to control the spread of Ebola
include surveillance, placement of suspected cases in quarantine 
for three weeks (the maximum estimated length of the incubation
period), education of hospital personnel and community members on the
use of strict barrier nursing techniques (i.e protective clothing and 
equipment, patient management), and the rapid burial or cremation of patients 
who die from the disease \cite{WHO1}. Their net effect, in our model, is 
to reduce the transmission rate $\beta$ from 
$\beta_0$ to $\beta_1 < \beta_0$. In practice, the impact of the
intervention is not instantaneous. Between the time of the onset of
the intervention to the time of full compliance, the transmission rate
is assumed to decrease gradually from $\beta_0$ to $\beta_1$ according
to\\
$$
\beta(t) = \left \{ \begin{array}{ll}

\beta_0 & t<\tau \\
\beta_1+ (\beta_0 - \beta_1)e^{-q (t-\tau)} & t \ge \tau
\end{array} \right.
$$

\noindent where $\tau$ is the time at which interventions start and
$q$ controls the rate of the transition from $\beta_0$ to
$\beta_1$. Another interpretation of
the parameter $q$ can be given in terms of $t_h = \frac{ln(2)}{q}$, 
the time to achieve $\beta(t) = \frac{\beta_0 + \beta_1}{2}$.

\subsection{Epidemiological data}

The data for the Congo (1995) and Uganda (2000) Ebola hemorrhagic
fever outbreaks include the identification dates of
the causative agent and data sources. The reported data are ($t_i$,
$y_i$), $i=1,...,n$ where $t_i$ denotes the $i^{th}$ reporting time
and $y_i$ the cumulative number of infectious cases from the beginning of the
outbreak to time $t_i$.\\

\begin{figure}[h*]
   \begin{center}
   \scalebox{0.5}{\includegraphics{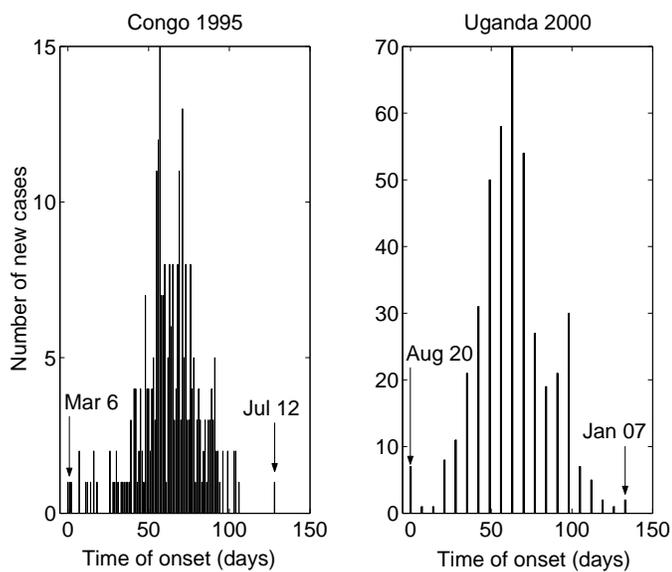}}
   \end{center}
   \caption{On the left, we have the \textit{daily} number of cases by date of symptom onset
     during the Ebola outbreak in Congo 1995 (Mar $6$-Jul $12$). On the right,
     we have the \textit{weekly} number of cases by date of symptom 
     onset during the Ebola outbreak in Uganda $2000$ (Aug $20$-Jan
     $07$). Data has been taken from refs. \cite{Khan1, WHO2}.}
\label{figDailycases}
\end {figure}

\noindent \textbf{Congo 1995.} This outbreak began in the Bandundu region, primarily
in Kikwit, located on the banks of the Kwilu River. The first case
(January 6) involved a 42-year old male charcoal worker and farmer
who died on January 13. The Ebola virus was not 
identified as the causative agent until May $9$. At that time, an international team 
implemented a control plan that involved active
surveillance (identification of cases) and education
programs for infected people and their family members. Family members
were visited for up to three weeks (maximum incubation period) after
their last identified contact 
with a probable case. Nosocomial transmission occurred in Kikwit General Hospital but it 
was halted through the institution of strict barrier nursing 
techniques that included the use of protective equipment and 
special isolation wards. A total of $315$ cases of Ebola 
were identified ($81\%$ case fatality). Daily Ebola cases by date of 
symptom onset from March $1$ through
July $12$ are available (Figure \ref{figDailycases}) \cite{Khan1}.\\

\noindent \textbf{Uganda 2000.} A total of $425$ cases ($53\%$ case
fatality) of Ebola were identified in three
districts of Uganda: Gulu, Masindi and Mbara. The onset of symptoms for
the first reported case was on August $30$, but the cause was
not identified as Ebola until October $15$ by the National Institute of
Virology in Johannesburg (South Africa). Active 
surveillance started during the third week of October. A plan that
included the voluntary hospitalization of probable cases 
was then put in place. Suspected cases were closely followed 
for up to three weeks. Other control measures included community
education (avoiding crowd gatherings 
during burials) and the systematic implementation of 
protective measures by health care personnel and 
the use of special isolation wards in hospitals. Weekly Ebola cases 
by date of symptom onset are available from the WHO (World Health
Organization) \cite{WHO2} (from August $20$, $2000$ through January $7$, $2001$) 
(Figure \ref{figDailycases}).\\

\subsection{Parameter Estimation}

Empirical studies in Congo suggest that the
incubation period is less than $21$ days with a mean of $6.3$ days
\cite{Breman1} and the infectious period is between $3.5$ and $10.7$
days. The model parameters $\Theta =(\beta_0$, $\beta_1$, $k$, $q$, $\gamma$) are
fitted to the Congo (1995) and Uganda (2000) Ebola outbreak data by 
\textit{least squares} fit to the cumulative number of cases $C(t,\Theta)$ in
eqn. (\ref{eqn1}). We used a computer program 
(Berkeley Madonna, Berkeley, CA) and appropriate initial
conditions for the parameters ($0<\beta<1$, $0<q<100$, $1<1/k<21$
\cite{Breman1}, $3.5<1/\gamma<10.7$ \cite{Piot1}). The optimization
process was repeated $10$ times (each time the program is fed with two
different initial conditions for each parameter) before the ``best fit'' was
chosen. The asymptotic variance-covariance $AV(\hat{\theta})$ of 
the least-squares estimate is\\
$$AV(\hat{\theta}) = \sigma^2 (\sum^n_{i=1} \nabla C(t_i, \Theta_0) \nabla C(t_i,\Theta_0)^{T})^{-1}$$ \\
\noindent which we estimate by \\
$$\hat{\sigma}^2 (\sum_{i=1}^n \hat{\nabla C}(t_i,\hat{\Theta}) \hat{\nabla
  C}(t_i,\hat{\Theta})^{T})^{-1}$$\\
\noindent where $n$ is the total number of observations, 
$\hat{\sigma}^2 = \frac{1}{n-5} \sum(y_i - C(t_i, \hat{\Theta}))^2$ 
and $\hat{\nabla C}$ are numerical derivatives of $C$.\\

\noindent For small samples, the confidence intervals based on these
variance estimates may not have the nominal coverage probability. For
example, for the case of Zaire $1995$, the $95 \%$ confidence interval for $q$ based on
asymptomatic normality is ($-0.26, 2.22$). It should be obvious that this interval
is not ``sharp'' as it covers negative values whereas we know $q \ge
0$. The likelihood ratio provides an attractive alternative to build
confidence sets (Figure \ref{figq}). Formally, these sets are of the form\\
$$\left \{ \Theta: \frac{\sum(y_i - C(t_i,\Theta))^2}{\sum(y_i - C(t_i,\hat{\Theta}))^2} \le A_{\alpha} \right \}$$ \\
\noindent where $A_{\alpha}$ is the $1-\alpha$ quantile of an $F$ distribution with
    appropriate degrees of freedom. Parameter estimates are given in
    Table \ref{TableParameters}.

\subsection{The Reproductive Number}

\noindent The basic reproductive number $R_0$ measures the average number 
of secondary cases generated by a primary case in a pool of mostly 
susceptible individuals \cite{AM,BC} and is an estimate of the
epidemic growth at the start of an outbreak if everyone is susceptible. That is, a primary case 
generates $R_0 = \frac{\beta_0}{\gamma}$ new cases on the average where $\beta_0$ is the
pre-interventions transmission rate and $1/\gamma$ is the mean
infectious period. The effective reproductive number
at time $t$, $R_{eff}(t) = \frac{\beta(t)}{\gamma} x(t)$, measures the average number of 
secondary cases per infectious
case $t$ time units after the introduction of the initial
infections and $x(t) = \frac{S(t)}{N} \approx 1$ as the population size is much larger than the resulting size of the outbreak (Table \ref{TableOutbreaks}). Hence, $R_{eff}(0)=R_0$. In a closed population, 
the effective reproductive number $R_{eff}(t)$ is
non-increasing as the size of the susceptible population
decreases. The case $R_{eff}(t) \le 1$ is of special interest as it highlights
the crossing of the threshold to eventual control of the outbreak. 
An intervention is judged successful if it reduces the effective 
reproductive number to a value less than one. In our model, the 
post-intevention reproductive number  $R_p = \frac{\beta_1}{\gamma}$ where $\beta_1$ denotes 
the post-intervention transmission rate. In general, the smaller $\beta_1$, the faster
an outbreak is extinguished. By the delta method \cite{Bickel1}, the variance of the estimated
basic reproductive number $\hat{R_0}$ is approximately\\

$$ V(\hat{R_0}) \approx \hat{R_0}^2 \ \{ \frac{V(\hat{\beta_0})}{\hat{\beta_0}^2} +
  \frac{V(\hat{\gamma})}{\hat{\gamma}^2} - \frac{2
    Cov(\hat{\beta_0}, \hat{\gamma})}{\hat{\beta_0}
    \hat{\gamma}} \}. $$

\subsection {The Effective Population Size}

\noindent A rough estimate of the population size in the Bandundu
region of Congo (where the epidemic developed) in 1995 is computed 
from the population size of the Bandundu region in $1984$
\cite{WGazzetter1} and annual population growth
 rates \cite{unhabitat1} (Table \ref{TableOutbreaks}). For the case
 of Uganda (2000), we adjusted the population sizes of the 
districts of Gulu, Masindi and Mbara in $1991$ and 
annual population growth rates \cite{UBOS1} (Table
\ref{TableOutbreaks}). These estimates are an upper bound of the
effective population size (those at risk of becoming infected) for
each region. Estimates of the effective population
size are essential when the incidence is modeled with the pseudo mass-action
assumption ($\beta(t) S I$) which implies that transmission grows linearly with the
population size and hence the basic reproductive number $R_0 (N) = \beta_0 N
/\gamma$. In our model, we use the true mass-action assumption
($\beta(t) S I/N$) which makes the model parameters (homogeneous
system of order $1$) independent of $N$ and hence the basic reproductive number can be
estimated by $R_0 = \beta_0 / \gamma$ \cite{CVF}. In fact, comparisons between
the pseudo mass-action and the true mass-action assumptions with
experimental data have concluded in favor of the later
\cite{JDH}. The model assumption that $N$ is constant is not critical as the
outbreaks resulted in a small number of cases compared to the size of the
population.

\subsection{Uncertainty Analysis on $R_0$}
Log-normal distributions seem to model well the incubation period distributions for
a large number of diseases \cite{Sartwell1}. Here, a log-normal
distribution is assumed for the incubation period of Ebola in our
uncertainty analysis. Log-normal distribution parameters are set from empirical
observations (mean incubation period is $6.3$ and the $95\%$ quantile
is $21$ days \cite{Breman1}). The infectious period is assumed to be uniformly
distributed in the range ($3.5-10.7$) days \cite{Piot1}. \\

\begin{figure}[h*]
   \begin{center}
   \scalebox{0.45}{\includegraphics{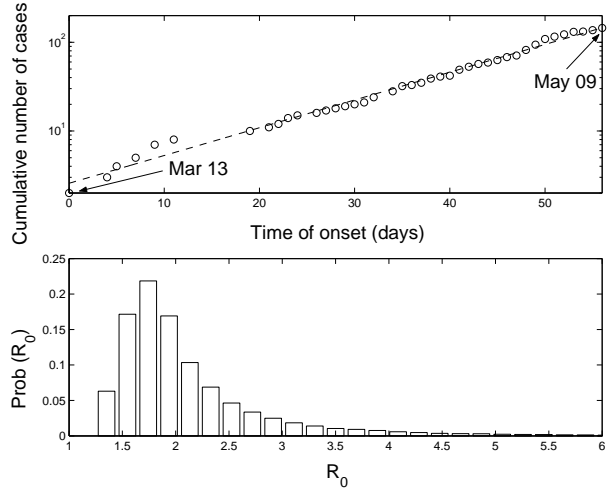}}
   \end{center}
   \caption{(Top) cumulative number of cases (log-lin scale) during the
     exponential growth phase of the Congo $1995$ epidemic as
     identified by the date of start of interventions ($09$ May
     $1995$ \cite{Khan1}). The model-free initial growth
     rate of the number of cases for Congo 1995 is $0.07$ (linear
     regression); (bottom)
     estimated distribution of $R_0$ from our uncertainty
     analysis (see text). $R_0$ lies in the
     interquartile range (IQR) ($1.66-2.28$) with a median of
     $1.89$. Notice that $100\%$ of the weight lies above $R_0=1$.}
\label{figR0uncertainty}
\end {figure}

\noindent A formula for the basic reproductive number $R_0$ that depends on the initial per-capita rate of
growth $r$ in the number of cases (Figure \ref{figR0uncertainty}), the incubation period 
($1/k$) and the infectious period ($1/\gamma$) can be obtained by linearizing 
equations $\dot{E}$ and $\dot{I}$ of system (\ref{eqn1}) around the disease-free equilibrium 
with $S=N$. The corresponding Jacobian matrix is given by:\\
\[
J=\left(\begin{array}{cc}
-k & \beta \\ 
k  & -\gamma \\
\end{array}\right),
\]

\noindent and the characteristic equation is given by:
\[
 r^2 + (k+\gamma) r + (\gamma - \beta) k = 0
\]

\noindent where the early-time and per-capita free growth $r$ is essentially the dominant eigenvalue. By solving for $\beta$ in terms of $r$, $k$ and $\gamma$, one can
obtain the following expression for $R_0$ using the fact that $R_0 = \beta/\gamma$:\\
$$ {R_0} = 1 + \frac{r^2 + (k+\gamma) r}{k \gamma}.$$ \\
\noindent Our estimate of the initial rate of growth $r$ for the Congo 1995 epidemic
is $r=0.07$ day$^{-1}$, obtained from the time series $y(t)$, $t<\tau$ of the
cumulative number of cases and assuming exponential growth ($y(t)
\propto e^{rt}$). The distribution of $R_0$ (Figure \ref{figR0uncertainty}) 
lies in the interquartile range (IQR) ($1.66-2.28$) with a median of $1.89$, generated from 
Monte Carlo sampling of size $10^5$ from the distributed epidemic
parameters ($1/k$ and $1/\gamma$) for fixed $r$ \cite{Blower1}. We give the median of $R_0$ (not the mean) as the resulting distribution of $R_0$ from our uncertainty analysis is skewed to the right.

\section{Results}

\begin{figure}[h*]
   \begin{center}
   \scalebox{0.45}{\includegraphics{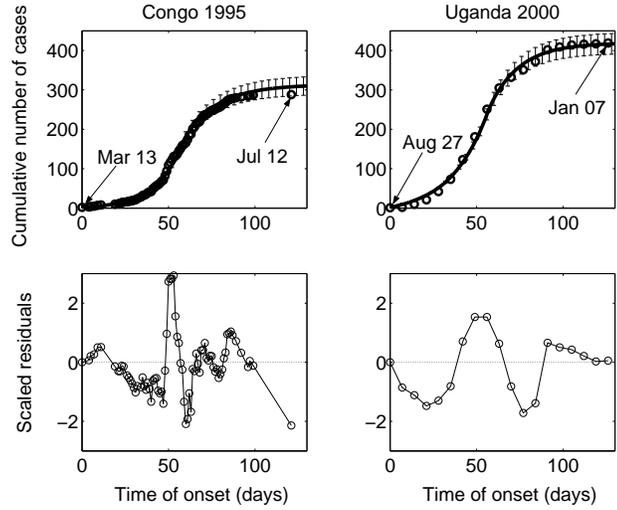}}
   \end{center}
   \caption{(Top) Comparison of the cumulative number of Ebola cases 
     during the Congo 1995 and Uganda 2000 Ebola outbreaks, as a function of the
     time of onset of symptoms. Circles are the data. The solid line
     is the average of $250$ Monte Carlo
     replicates and the error bars represent the standard
     error around the mean from the simulation
     replicates using our parameter estimates (Table
     \ref{TableParameters}). For the case of Congo $1995$, simulations were begun on $13$
     Mar $1995$. A reduction in the transmission rate $\beta$ due to
     the implementation of interventions occurs on $09$ May 1995 (day $56$)
     \cite{Khan1}. For the case of Uganda $2000$, simulations start on
     $27$ August  $2000$ and interventions take place on $22$
     October $2000$ (day $56$) \cite{WHO2}; (bottom) comparison of the residuals
     (difference between the data and the model best fit) scaled by
     the standard deviation for the cases of Congo and Uganda.}
\label{figmodel2}
\end {figure}

\begin{figure}[h*]
   \begin{center}
   \scalebox{0.45}{\includegraphics{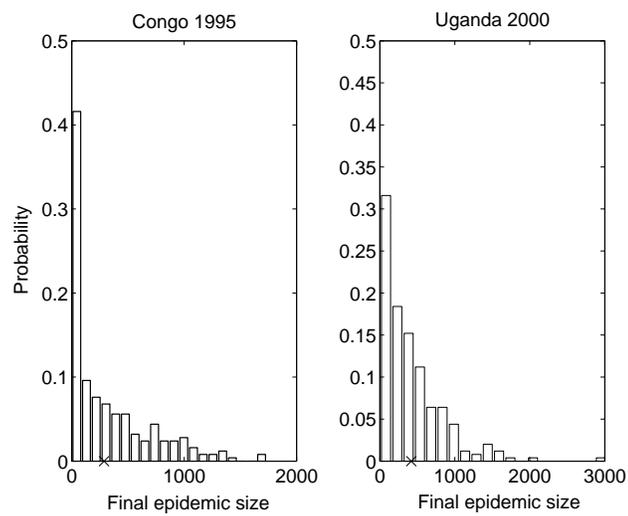}}
   \end{center}
   \caption{The final epidemic size distributions for
     the cases of Congo $1995$ and Uganda $2000$ obtained from $250$
     Monte Carlo replicas. Crosses (X) represent the final
     epidemic size from data.}
\label{figOutbreaksizedistr}
\end {figure}

\begin{figure}[h*]
   \begin{center}
   \scalebox{0.45}{\includegraphics{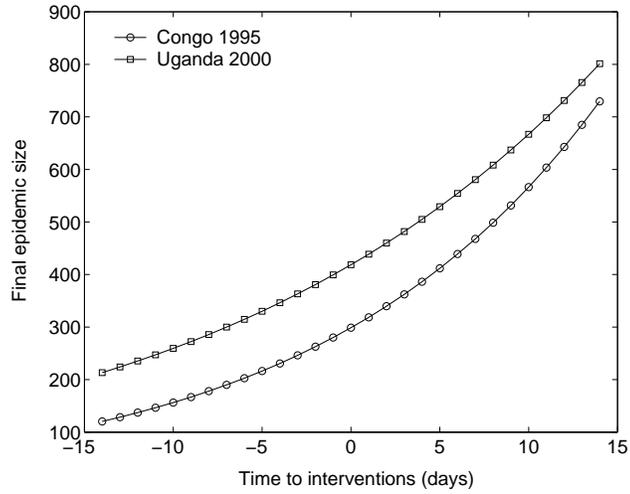}}
   \end{center}
   \caption{Sensitivity of the final epidemic size to the time of
     start of interventions. Here negative numbers represent 
     number of days before the actual reported intervention date (Table
     \ref{TableOutbreaks}) and positive numbers represent a delay
     after the actual reported intervention date ($\tau=0$). All other
     parameters have been fixed to their baseline values (Table
     \ref{TableOutbreaks}). The final epidemic size grows
     exponentially as expected with the time of interventions
     with a rate of $0.06$ for the case of Congo and $0.05$ for the
     case of Uganda.}
\label{figSensInterv}
\end {figure}

\indent Using our parameter estimates (Table \ref{TableParameters}), we
estimate an $R_0$ of $1.83$ (SD $0.06$) for Congo (1995) and
$1.34$ (SD $0.03$) for Uganda (2000). 
\noindent The effectiveness of interventions is often quantified in terms of the
reproductive number $R_p$ after interventions are put in place. For
the case of Congo $R_p = 0.51$ (SD $0.04$) and $R_p = 0.66$ (SD
$0.02$) for Uganda allowing us to conclude that in both cases, the intervention
was successful in controlling the epidemic. Furthermore, the time to
achieve a transmission rate of $\frac{\beta_0 + \beta_1}{2}$ ($t_h$)
is $0.71$ ($95\%$ CI ($0.02, 1.39$)) days and $0.11$ ($95 \%$ CI ($0, 0.87$)) days
for the cases of Congo and Uganda respectively after the time at which interventions begin.\\
\noindent We use the estimated parameters to simulate the Ebola outbreaks in Congo (1995)
and Uganda (2000) via Monte Carlo simulations of the stochastic model of Section $2.1$ \cite{Renshaw1}. 
There is very good agreement between the mean of the stochastic
simulations and the reported cases despite the
``wiggle'' captured in the residuals around the time $\tau$ of the 
start of interventions (Figure \ref{figmodel2}). The
empirical distribution of the final epidemic sizes for the cases of Congo
$1995$ and Uganda $2000$ are given in Figure \ref{figOutbreaksizedistr}.\\
\noindent The final epidemic size is sensitive to
the start time of interventions $\tau$. Numerical solutions
(deterministic model) show that the final epidemic
size grows exponentially fast with the initial time of
interventions (not surprising as the intial epidemic growth is driven by
exponential dynamics). For instance, for the case of
Congo, our model predicts that there would have been $20$ more cases
if interventions had started one day later (Figure \ref{figSensInterv}). 

\section{Discussion}

\indent Using epidemic-curve data from two major Ebola hemorrhagic fever
outbreaks \cite{Khan1, WHO2}, we have estimated the basic reproductive
number ($R_0$) (Table \ref{TableOutbreaks}). Our estimate of $R_0$ (median is $1.89$)
obtained from an uncertainty analysis \cite{Blower1} by simple random sampling (Figure
\ref{figR0uncertainty}) of the parameters $k$ and $\gamma$ distributed 
according to empirical data from the Zaire (now the Democratic Republic of Congo) $1976$ Ebola outbreak \cite{Breman1, Piot1} is in agreement with our estimate of $R_0 = 1.83$ from the outbreak in Congo $1995$ (obtained from least squares fitting 
of our model (\ref{eqn1}) to epidemic curve data). \\
The difference in the basic reproductive
numbers $R_0$ between Congo and Uganda is due to our different estimates for the
infectious period ($1/\gamma$) observed in these two places.  Their
transmission rates $\beta_0$ are quite similar (Table \ref{TableParameters}). Our
estimate for the infectious period for the case of Congo ($5.61$ days)
is slightly larger than that of Uganda ($3.50$ days). Clearly, a larger infectious
period increases the likelihood of infecting a susceptible
individual and hence increases the basic reproductive number. 
The difference in the infectious periods might be due to differences 
in virus subtypes \cite{Niikura1}. The Congo outbreak was caused by the Ebola-Zaire
virus subtype \cite{Khan1} while the Uganda outbreak was caused by 
the Ebola-Sudan virus subtype \cite{WHO2}.\\
\noindent The significant reduction from the basic reproductive number ($R_0$) to the post-intervention reproductive number ($R_p$) in our estimates for Congo and Uganda shows that the implementation of control measures such as education, contact tracing and quarantine will have a significant effect on lowering the effective reproductive rate of Ebola. 
Furthermore, estimates for the time to achieve 
$\frac{\beta_0 + \beta_1}{2}$ have been provided (Table \ref{TableParameters}).\\
\noindent We have explored the sensitivity of the final epidemic size to the
starting time of interventions. The exponential increase of the final
epidemic size with the time of start of interventions (Figure
\ref{figSensInterv}) supports the idea that the rapid
implementation of control measures should be considered as a critical
component in any contingency plan against disease outbreaks specially
for those like Ebola and SARS for which no specific treatment or
vaccine exists. A  two-week delay in implementing public health
measures results in an approximated doubling of the final outbreak
size. Because the existing control measures cut the transmission rate to 
less than half, we should seek and support further improvement in the effectiveness of 
interventions for Ebola. A mathematical model that considers basic public health interventions for SARS control in Toronto supports this conclusion \cite{Chowell1, Chowell2}. Moreover, computer simulations show that small perturbations to the rate $q$ at which interventions are put fully in place do not have a significant effect on the final epidemic size. The rapid identification of an outbreak, of course, remains the strongest determinant of the final outbreak size.\\
\noindent Field studies of Ebola virus are difficult to conduct due to
the high risk imposed on the scientific and medical personnel
\cite{Nature2}. Recently, a new vaccine that makes use of an
\textit{adenovirus technology} has been shown to give cynomolgus macaques
protection within $4$ weeks of a single jab \cite{Nature3,
  Nature4}. If the vaccine turns out to be effective in humans, then
its value should be tested. A key question would be ``What are the 
conditions for a successful target vaccination campaign during an Ebola outbreak?'' 
To address questions of this type elaborate models need to be developed.

\newpage

\section*{Tables \& Figures}

\renewcommand{\thefootnote}{\fnsymbol{footnote}}

\begin{table}[h*]
\begin {tabular}{||l|l|c|c|c|c||}
\hline
& & \multicolumn{2}{c|}{Congo 1995}
& \multicolumn{2}{c||}{Uganda 2000} \\
\hline
Parameter & Definition & Estim. & S.\ D. & Estim. & S. \ D. \\
\hline
$\beta_0$ & Pre-interventions transmission rate (days$^{-1}$) & $0.33$ & $0.06$ &
$0.38$ & $0.24$\\
$\beta_1$ & Post-interventions transmission rate (days$^{-1}$) & $0.09$ & $0.01$ &
$0.19$ & $0.13$\\
$t_h$ & Time to achieve $\frac{\beta_0 + \beta_1}{2}$ (days) & $0.71$ & $(0.02, 1.39)$\footnotemark[2] &
$0.11$ & $(0, 0.87)$\footnotemark[2]\\
$1/k$ & Mean incubation period (days) & $5.30$ & $0.23$ & $3.35$ & $0.49$ \\
$1/\gamma$ & Mean infectious period (days) & $5.61$ & $0.19$ & $3.50$ & $0.67$\\
\hline
\end{tabular}
\caption{Parameter definitions and baseline estimates obtained
from the best fit of the model equations (\ref{eqn1}) to the 
epidemic-curve data of the Congo $1995$ and Uganda $2000$
outbreaks (Figure \ref{figmodel2}). The parameters  were optimized
by a computer program (Berkeley Madonna, Berkeley, CA) using a \textit{least
squares} fitting technique and appropriate initial
conditions for the parameters ($0<\beta<1$, $0<q<100$, $1<1/k<21$
\cite{Breman1}, $3.5<1/\gamma<10.7$ \cite{Piot1}). The optimization
process was repeated $10$ times (each time the program is fed with two
different initial conditions for each parameter) before the ``best fit'' was
chosen.}
\label{TableParameters}
\end{table}
\vfill

\begin{figure}[h*]
   \begin{center}
   \scalebox{0.45}{\includegraphics{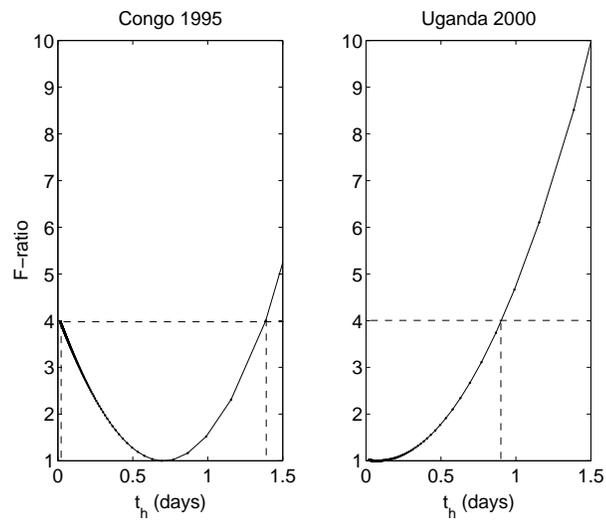}}
   \end{center}
   \caption{$95 \%$ confidence intervals for $t_h$ ($t_h =
     \frac{log(2)}{q}$), the time to achieve a transmission rate of
     $\frac{\beta_0 + \beta_1}{2}$, obtained
     from the likelihood ratio as described in the text.}
\label{figq}
\end {figure}

\footnotetext[2]{95 $\%$ CI (Figure \ref{figq}).}

\begin{table}[h*]
\begin {tabular}{||c|c|l|c|c|c|c|c||}
\hline
Outbreak & Eff. Pop. Size (N) & Start of interv. & Fatality rate
($\%$) & Estim. $R_0$ & S.D. $R_0$ \\
\hline
 Congo 1995 & $5,364,500$\footnotemark[1] & May $9$, $1995$ \cite{Khan1} & $81\%$
 \cite{Khan1} & $1.83$ & $0.06$ \\
 Uganda 2000 & $1,867,200$ \footnotemark[5] & Oct $22$, $2000$ \cite{WHO2} &
 $53\%$ \cite{WHO2} & $1.34$ & $0.03$ \\
\hline
\end{tabular}
\caption{Population parameters and estimated $R_0$ for the Congo $1995$
  and the Uganda $2000$ Ebola outbreaks. Notice that even though our expression for $R_0$ is independent of $N$, our model is not independent of $N$ and hence the corresponding population sizes for Congo and Uganda are used in the least-squares estimation of the parameters.}
\label{TableOutbreaks}
\end{table}

\footnotetext[1]{Adjusted from population size of the Bandundu region
 in $1984$ \cite{WGazzetter1}  using the annual population growth
 rates \cite{unhabitat1}.}

\footnotetext[5]{Adjusted from the population sizes of the districts of Gulu,
 Masindi and Mbara (where the outbreak developed) in $1991$ using the 
annual population growth rates \cite{UBOS1}.}

\vfill

\vfill

\end{document}